\documentclass[aps,pre,superscriptaddress,citeautoscript,twocolumn,reprint,longbibliography,floatfix]{revtex4-2}
\usepackage{graphicx,amssymb,amsmath,bm,cprotect,comment,physics}
\newcommand{\rd}{\mathrm{d}}
\newcommand{\expct}[1]{\langle{#1}\rangle}

\newcommand{\diff}[2]{\frac{\mathrm{d} #1}{\mathrm{d} #2}}

\renewcommand{\eqref}[1]{Eq.\,(\ref{#1})}
\newcommand{\eqsref}[1]{Eqs.\,(\ref{#1})}
\newcommand{\pref}[1]{(\ref{#1})}
\newcommand{\figref}[1]{Fig.\,\ref{#1}}

\newcommand{\secref}[1]{Sec.\,\ref{#1}}
\newcommand{\secsref}[1]{Secs.\,\ref{#1}}

\begin{document}

\title{Population dynamics simulations of large deviations for three subclasses of the Kardar-Parisi-Zhang universality class}
%Large deviation  using Population dynamics simulation

\author{Yuta Yanagibashi}
\affiliation{Department of Physics,\! The University of Tokyo,\! 7-3-1 Hongo,\! Bunkyo-ku,\! Tokyo 113-0033,\! Japan}%
%\affiliation{Department of Physics, The University of Tokyo, Tokyo, Japan}%

\author{Kazumasa A. Takeuchi}
\email{kat@kaztake.org}
\affiliation{Department of Physics,\! The University of Tokyo,\! 7-3-1 Hongo,\! Bunkyo-ku,\! Tokyo 113-0033,\! Japan}%
\affiliation{Institute for Physics of Intelligence,\! The University of Tokyo,\! 7-3-1 Hongo,\! Bunkyo-ku,\! Tokyo 113-0033,\! Japan}%
\affiliation{Universal Biology Institute,\! The University of Tokyo,\! 7-3-1 Hongo,\! Bunkyo-ku,\! Tokyo 113-0033,\! Japan}%
%\affiliation{Department of Physics, The University of Tokyo, Tokyo, Japan}%

\date{\today}

\begin{abstract}
Recent theoretical studies have gradually deepened our understanding of the one-dimensional (1D) Kardar-Parisi-Zhang (KPZ) universality class even in the large deviation regime, but numerical methods for studying KPZ large deviations remain limited.
Here we implement a method based on the population dynamics algorithm for studying large deviations of time-integrated local currents in the totally asymmetric simple exclusion process (TASEP), which is a pragmatic model in the 1D KPZ class. 
Carrying out simulations for the three representative initial conditions, namely step, flat, and stationary ones, we not only confirm theoretical predictions available for the step case, but also characterize large deviations for the flat and stationary cases which have not been investigated before.
We reveal in particular an unexpected robustness of the deeply negative large deviation regime with respect to different initial conditions. 
We attribute this robustness to the spontaneous formation of a wedge shape in interface profile.
Our population dynamics approach may serve as a versatile method for studying large deviations in the KPZ class numerically and, potentially, even experimentally.
\end{abstract}

\maketitle

\section{Introduction}
%%KPZnogairyaku
The importance of the Kardar-Parisi-Zhang (KPZ) universality class \cite{Kardar.etal-PRL1986} for many-body dynamical processes is widely acknowledged (for reviews, see, e.g, \cite{Corwin-RMTA2012,Takeuchi-PA2018}).
First, it describes a surprisingly broad range of phenomena, starting from interface growth, stochastic particle transport, and directed polymers. 
Most recently, even polariton condensates \cite{Fontaine.etal-N2022}, time crystals \cite{Daviet.etal-PRL2025}, and integrable spin chains \cite{Gopalakrishnan.Vasseur-ARCMP2024,Takeuchi.etal-PRL2025} are in the scope.
Second, the KPZ class in one dimension (1D) has been thoroughly studied through exact solutions \cite{Corwin-RMTA2012,Takeuchi-PA2018}, revealing a wealth of nontrivial universal statistical properties.
For example, it has been shown that the distribution function of typical fluctuations is given by a family of Tracy-Widom distributions, originally formulated in random matrix theory, and that it depends on the choice of the initial condition.
The following three cases, sometimes called the universality subclasses, are particularly important: the Tracy-Widom distribution for the Gaussian unitary ensemble (GUE) appearing for the step initial condition, that for the Gaussian orthogonal ensemble (GOE) for the flat initial condition, and the Baik-Rains distribution for the stationary initial condition.
In the literature, these universal laws have been occasionally compared even with the central limit theorem \cite{Corwin-NAMS2016}.

Similarly to the central limit theorem, the Tracy-Widom laws for the 1D KPZ class govern the distribution of \textit{typical} fluctuations, leaving \textit{large deviations} \cite{[{See, e.g., }][{ for a general review on large deviations.}]Touchette-PR2009} beyond the scope.
Here, typical fluctuations refer to fluctuations growing with time as $t^{1/3}$, governed by the growth exponent of the 1D KPZ class, while large deviations are those growing in the order of $t$.
Even though large deviations are generally less universal than typical fluctuations, recent theoertical and mathematical studies have unveiled interesting statistical properties that may be partly shared in the KPZ class (see, e.g., \cite{LeDoussal.etal-EPL2016,Sasorov.etal-JSM2017,Corwin.etal-PRL2018,Krajenbrink-thesis2020,Quastel.Tsai-CPAM2024,Das.etal-a2024,Krajenbrink.LeDoussal-a2025}).
For example, the probability density $P(\delta h,t)$ of local interface height fluctuations $\delta h$ at large time $t$ can be classified to the following three regimes, with asymmetric time-dependence between the negative and positive tails \cite{LeDoussal.etal-EPL2016}:
\begin{equation}
  P(\delta h,t) \sim \begin{cases}
    e^{-t^2 \Phi_-^{(+)}(\delta h/t)}, & \delta h \sim -\mathcal{O}(t) < 0, \\
    t^{-1/3} f^{(+)}(\delta h/t^{1/3}), & \delta h \sim \mathcal{O}(t^{1/3}), \\
    e^{-t \Phi_+^{(+)}(\delta h/t)}, & \delta h \sim \mathcal{O}(t) > 0,
  \end{cases}
\end{equation}
if the KPZ nonlinearity takes a positive coefficient.
Otherwise, i.e., for negative KPZ nonlinearity,
\begin{equation}
  P(\delta h,t) \sim \begin{cases}
    e^{-t \Phi_-^{(-)}(\delta h/t)}, & \delta h \sim -\mathcal{O}(t) < 0, \\
    t^{-1/3} f^{(-)}(\delta h/t^{1/3}), & \delta h \sim \mathcal{O}(t^{1/3}), \\
    e^{-t^2 \Phi_+^{(-)}(\delta h/t)}, & \delta h \sim \mathcal{O}(t) > 0.
  \end{cases}  \label{eq:P3regimes}
\end{equation}
%Note that \eqref{eq:P3regimes} is for the case of the positive KPZ nonlinearity, which the plus signs of the superscripts indicate.
%For the negative KPZ nonlinearity, the positive and negative tails are exchanged, so that $P(\delta h,t) \sim e^{-t\Phi_-^{(-)}(\delta h/t)}$ for $\delta h \sim -\mathcal{O}(t) < 0$ and $P(\delta h,t) \sim e^{-t^2\Phi_+^{(-)}(\delta h/t)}$ for $\delta h \sim \mathcal{O}(t) < 0$.
Here we assumed that $\delta h$ and $t$ are appropriately rescaled and dimensionless.
%, and the sign of $\delta h$ is chosen such that it corresponds to the case of the positive nonlinear coefficient of the KPZ equation. 
%For negative nonlinearity, the sign of $\delta h$ is set to be opposite to that of the bare interface height.
In the following, we may omit the superscript in $\Phi_\pm^{(\pm)}$ and $f^{(\pm)}$ indicating the sign of the KPZ nonlinearity, unless it has to be specified explicitly.

The quantities of interest are the rate functions $\Phi_\pm(\cdot)$ or equivalently their Legendre transforms.
While analytical solutions were given in the literature for some solvable models and specific initial conditions \cite{Johansson-CMP2000,LeDoussal.etal-EPL2016,Sasorov.etal-JSM2017,Krajenbrink-thesis2020}, exact results on large deviations, in particular those for late times and on the initial condition dependence, are considerably limited compared with typical fluctuations.
Experimentally, while exact results for typical fluctuations were largely tested and confirmed in liquid crystal turbulence \cite{Takeuchi.Sano-PRL2010,Takeuchi.etal-SR2011,Takeuchi.Sano-JSP2012,Fukai.Takeuchi-PRL2020,Iwatsuka.etal-PRL2020,Takeuchi-PA2018}, large deviations remain beyond the scope, almost by definition, because large deviations concern rare events that do not occur in realistic time scales.

\begin{figure}
  \centering
  \includegraphics{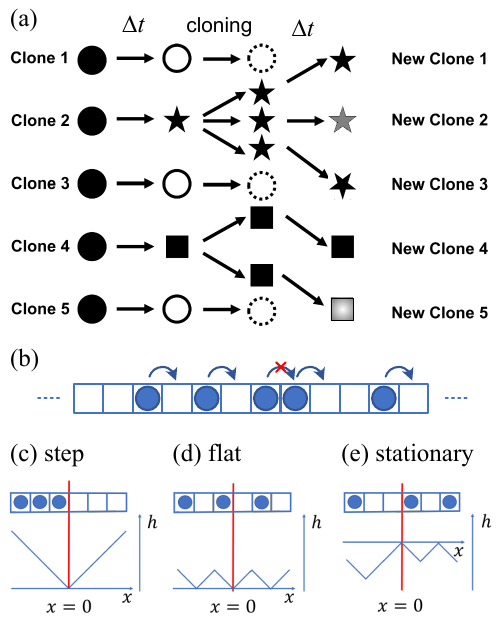}
  \caption{
  Model and method. 
  (a) Concept of the population dynamics algorithm. 
  (b) The model TASEP. Particles hop stochastically to their right neighbor sites at a constant rate if the target sites are empty.
  (c)-(e) Three initial conditions considered in this work. The top and bottom panels show particle configurations and the corresponding interface profiles, respectively.
  }
  \label{fig:model}
\end{figure}

Numerical approaches are therefore expected to play an important role.
Numerically, various algorithms have been devised to efficiently sample rare events, known by the name of importance sampling \cite{Touchette-PR2009}. 
Such a sampling is typically achieved by adjusting statistical weights so that those events are no longer rare, by modifying the time evolution rule for example, then converting the biased statistical quantities to the original, unbiased ones.
This strategy was successfully applied to local heights of the KPZ equation  \cite{Hartmann.etal-EL2018,Hartmann.etal-PRE2020}, but the initial condition dependence of large deviations at late times is yet to be investigated.
More importantly, since modifying the time evolution rule is impossible in experiments, it is desired to develop another method that may in principle be implemented in experiments too.
In this context, it is interesting to test another type of importance sampling that does not modify the time evolution rule, namely the population dynamics (or cloning) method \cite{Touchette-PR2009}, which replicates or kills trajectories during the time evolution to assign biased statistical weights to them (\figref{fig:model}(a)).
In the KPZ literature, this method was applied to global heights \cite{Giardina.etal-PRL2006,Giardina.etal-JSP2011} but not to local heights. 
The latter is of interest because of the characteristic time dependence of \eqref{eq:P3regimes} and other predictions available from exact studies.

Here we report population dynamics simulations for studying large deviations of local heights, with a particular focus on their initial condition dependence.
Specifically, we use the totally asymmetric simple exclusion process (TASEP), one of the best studied models in mathematical approaches to KPZ \cite{Corwin-RMTA2012,Takeuchi-PA2018}.
After describing the model and the simulation method in \secref{sec:model}, we will present results for the three representative initial conditions of the KPZ class, namely the step, flat, and stationary initial conditions, in \secsref{sec:step}-\ref{sec:stationary}.
Concluding remarks are given in \secref{sec:conclusion}.

\section{Model and Method} \label{sec:model}

\subsection{Model}

The studied model, TASEP, is the strong asymmetry limit of the asymmetric simple exclusion process (ASEP) \cite{Derrida-PR1998,Golinelli.Mallick-JPA2006}, a paradigmatic model of stochastic particle transport on a one-dimensional lattice.
In TASEP, particles hop stochastically to their right neighbor sites at a constant rate (set to be $1$) if the target sites are empty (\figref{fig:model}(b)).
The only interaction between particles is volume exclusion, which prevents them from hopping to sites that are already occupied; otherwise particles undergo stochastic hopping independently.
This model is known (and mathematically proven) to be in the KPZ class.
Indeed, by replacing occupied and unoccupied sites with downward and upward slopes, respectively, one can map particle configurations to interface profiles (\figref{fig:model}(c)-(e)) and each hopping of a particle corresponds to a local growth of the interface at the corresponding position.
%Therefore, it is a model of interface growth with nonlinear interactions (due to the volume exclusion) and as such it is not surprising to be in the KPZ class.
The increment of the interface height $h(x,t)$ is given by the time-integrated current of particles, i.e., the total number of particles that have passed position $x$ rightward until time $t$.
In the following, we only analyze $h$ at $x=0$, which is the center of the lattice of size $N$ (see \figref{fig:model}(c)-(e) again).
With this, we define the rescaled height fluctuation $\delta h$ by
\begin{equation}
    \delta h \equiv \frac{h - \rho(1-\rho)t}{\rho(1-\rho)}  \label{eq:deltah}
\end{equation}
with particle density $\rho$ fixed to $1/2$ here.
%The sign of $\delta h$ is chosen so that it is consistent with \eqref{eq:P3regimes}, reflecting the negative KPZ nonlinearity of TASEP.
Note that TASEP is known to have negative KPZ nonlinearity \cite{Johansson-CMP2000,Corwin-RMTA2012,Takeuchi-PA2018}, so that the large deviation laws are given by \eqref{eq:P3regimes}.
The superscript of $\Phi_\pm^{(-)}$ and $f^{(-)}$ will be omitted.

In the present work, we study the three representative subclasses of the one-dimensional KPZ class, corresponding to the step, flat, and stationary initial conditions (\figref{fig:model}(c)-(e)).
The step initial condition is a configuration where the left half of the system is fully occupied and the right half is fully empty (\figref{fig:model}(c)), and typical fluctuations in this case are known to show the GUE Tracy-Widom distribution \cite{Johansson-CMP2000,Corwin-RMTA2012,Takeuchi-PA2018}.
The flat initial condition refers to a configuration where occupied and unoccupied sites are alternating (\figref{fig:model}(d)) and typical fluctuations show the GOE Tracy-Widom distribution.
For the stationary initial condition, an initial configuration is drawn from the Bernouilli distribution with density $\rho=1/2$ (\figref{fig:model}(e)), i.e., each site is set to be occupied or unoccupied with equal probability.
Here we also impose the constraint that the total number of occupied sites is half of the number of sites, $N$.
Practically, we choose $N/2$ out of $N$ sites randomly and put particles therein.
Typical fluctuations in this stationary subclass are known to be characterized by the Baik-Rains distribution.

\subsection{Population dynamics algorithm}

We adapted the population dynamics algorithm developed in Refs.\ \cite{Giardina.etal-PRL2006,Giardina.etal-JSP2011} to study large deviations of local height fluctuations $\delta h$ of TASEP. 
Essentially, we generate a given number (denoted by $N_\mathrm{cl}$) of ``clones'' starting from an identical initial condition. 
They evolve independently, but at a given interval, some are duplicated and others are killed, according to their value of $\delta h$ at $x=0$ and the bias parameter $k$ we set for each simulation.
For $k>0$ (resp. $k<0$), clones with larger (resp. smaller) $\delta h$ are replicated more at the expense of other clones, in such a way that the total number of clones is kept constant at $N_\mathrm{cl}$.
Specifically, we implement it as follows, for each time step of size $\Delta t=1$:
\begin{enumerate}
  \setlength{\parskip}{0cm} % 段落間
  \setlength{\itemsep}{0cm} % 項目間
  \item Each clone (index $i=1, 2, \cdots, N_\mathrm{cl}$) evolves during $\Delta t$ according to the time evolution rule of TASEP.
  \item For each clone $i$, the change of $\delta h$ during $\Delta t$, denoted by $\Delta h_i$, is recorded. Then the partition function $Z_t=\sum_{i=1}^{N_\mathrm{cl}} e^{k\Delta h_i}$ is evaluated.
  \item Each clone $i$ is replicated to $n_i=\lfloor\frac{e^{k\Delta h_i}}{Z_t}N_\mathrm{cl}+\eta \rfloor$ clones, with a random number $\eta \in [0,1)$ drawn from the uniform distribution. If $n_i=0$, the clone is eliminated. 
  \item Since the resulting number of clones may deviate slightly from $N_\mathrm{cl}$, we randomly select an appropriate number of clones to duplicate or eliminate, thereby precisely setting the total number of clones to $N_\mathrm{cl}$.
\end{enumerate}
Repeating these steps, we obtain a time series of the partition function, $\{Z_t\}$.
From this, we can evaluate the cumulant generating function (CGF) 
\begin{equation}
    \lambda(k,t) \equiv \ln \expct{e^{k\delta h}} = \ln \int e^{k\delta h} P(\delta h, t) \rd(\delta h)  \label{eq:lambdaDef}
\end{equation}
of the original (unbiased) TASEP, by
\begin{equation}
      \lambda(k,t) = \ln{\left(\prod_{t'=0}^{t} \frac{Z_{t'}}{N_\mathrm{cl}}\right)}. \label{eq:lambda_clone}
\end{equation}
%and the scaled CGF $\mu_\pm(k,t)$
By substituting the function form of each large deviation regime in \eqref{eq:P3regimes} for $P(\delta h,t)$ in \eqref{eq:lambdaDef}, we obtain, for large $t$,
%Corresponding to the negative and positive large deviation regimes in \eqref{eq:P3regimes}, we expect $\lambda(k,t)$ to behave as follows:
\begin{equation}
    \lambda(k,t) \simeq \begin{cases} t \mu_-(k), & (k<0), \\ t^2 \mu_+(k/t), & (k>0), \end{cases}  \label{eq:lambda_2regimes}
\end{equation}
with the scaled CGF $\mu_\pm(k)$ defined by the following Legendre transform:
\begin{equation}
    \mu_\pm(k) = \sup_z [kz - \Phi_\pm(z)].   \label{eq:Legendre_phi2mu}
\end{equation}

%\begin{gather}
%  \mu_-(k,t) = \frac{1}{t} \lambda(k,t), \quad(k<0), \label{eq:mu-} \\
%  \mu_+(k/t,t) = \frac{1}{t^2} \lambda(k,t), \quad(k>0). \label{eq:mu+}
%\end{gather}
%The scaled CGF defined thereby is expected to converge in $t \to \infty$.
%The converged value $\mu_\pm(k) \equiv \lim_{t\to\infty}\mu_\pm(k,t)$ is known to be related to the rate function $\Phi_\pm(z)$ by the following Legendre transforms:
%\begin{align}
%    \Phi_\pm(z) = \sup_k [kz - \mu_\pm(k)],   \label{eq:Legendre_mu2phi} \\
%    \mu_\pm(k) = \sup_z [kz - \Phi_\pm(z)],   \label{eq:Legendre_phi2mu}
%\end{align}

\begin{figure}
  \centering
  \includegraphics[keepaspectratio,width=8.6cm]{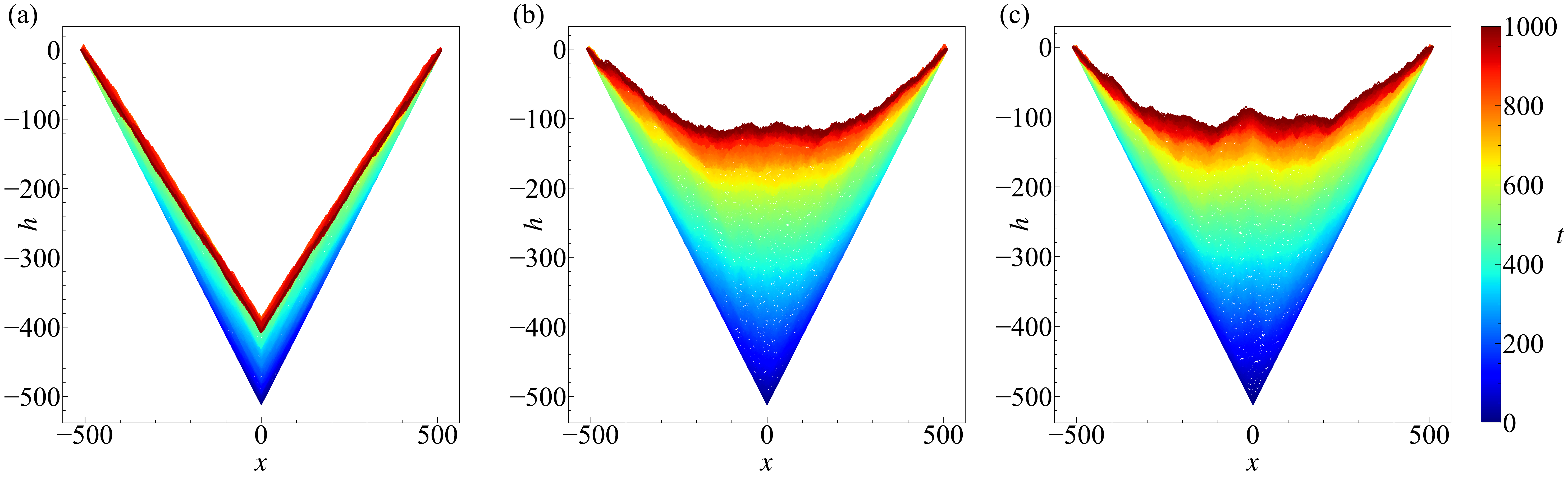}
  \caption{
  Examples of time evolution of interface profiles for the step initial condition, for $k=-1$ (a), $k=0$ (b), and $k=1$ (c). Simulation parameters: $N=1024, N_\mathrm{cl} = 10000$.
  }
  \label{fig:profile}
\end{figure}

Figure~\ref{fig:profile} shows time evolution of typical interface profiles for various bias parameters $k$, in the case of the step initial condition. If $k<0$, particles tend not to cross $x=0$, and as a result the wedge-shaped initial interface profile tends to persist (\figref{fig:profile}(a)).
If $k>0$, particles tend to cross $x=0$ as soon as they arrive at the site on its left, so that the interface is pinched upward at $x=0$ (\figref{fig:profile}(c)).
Note that the effect of bias is quite asymmetric.
Indeed, while for $k<0$ it is sufficient to prevent the particle right before $x=0$ from jumping to keep $\delta h$ small, for $k>0$, in order to bias $\delta h$ positively, we need to promote hopping not only to the particle right before $x=0$ but also to subsequent particles to make the site right before $x=0$ occupied as soon as possible.
Therefore, unlike the negative bias, the positive bias essentially involves multi-body effect, and this results in the negative-positive asymmetry of the large deviation laws in \eqref{eq:P3regimes}.

\section{Step initial condition} \label{sec:step}

\begin{figure*}
  \centering
  \includegraphics[keepaspectratio,width=\hsize]{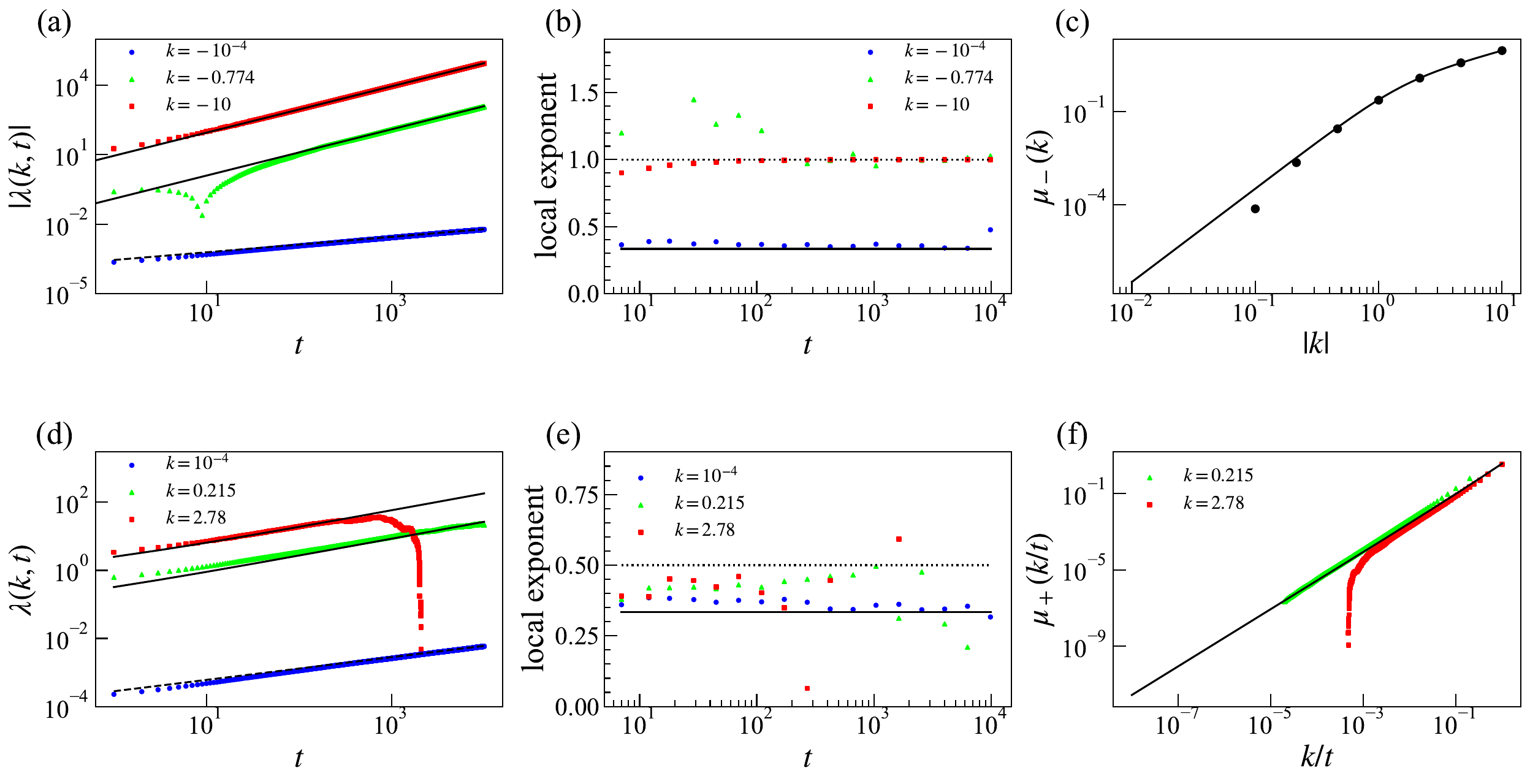}
  \caption{
    Simulation results for the step initial condition with $N=10000$ and $N_\mathrm{cl} = 10000$.
    (a)(b) The absolute value of CGF $|\lambda(k,t)|$ (a) and its local exponent $\diff{\ln |\lambda(k,t)|}{\ln t}$ (b) for negative bias $k<0$.
    In (a), simulation data (symbols) are compared with Johansson's exact solution for negative large deviations (solid lines), \eqref{eq:step_mu-}, or the prediction for typical fluctuations (dashed line), \eqref{eq:step_lambda_typical}.
    (c) The scaled CGF $\mu_-(k)$ for the negative tail.
    Simulation data of $\lambda(k,t)/t$ with $t=10000$ (symbols) are compared with the exact solution (line), \eqref{eq:step_mu-}.
    (d)(e) The CGF $\lambda(k,t)$ (d) and its local exponent $\diff{\ln \lambda(k,t)}{\ln t}$ (e) for positive bias $k>0$.
    In (d), simulation data (symbols) are compared with Johansson's exact solution for positive large deviations (solid lines), \eqref{eq:step_mu+}, or the prediction for typical fluctuations (dashed line), \eqref{eq:step_lambda_typical}.    
    (f) The scaled CGF $\mu_+(k/t)$ for the positive tail.
    Simulation data of $\lambda(k,t)/t^2$ (symbols) are plotted against $k/t$ and compared with the exact solution (line), \eqref{eq:step_mu+}.    
  }
  \label{fig:step}
\end{figure*}

We first study the case of the step initial condition (\figref{fig:model}(c)).
This serves as a testbed of our population dynamics method for local heights, because exact solutions are available for this case \cite{Johansson-CMP2000,LeDoussal.etal-EPL2016,1note}:
\begin{align}
  &\Phi^\mathrm{step}_-(z)=|z|^{1/2}-\frac{1}{2}(1+z)\ln{\frac{1+|z|^{1/2}}{1-|z|^{1/2}}}, \quad (-1<z<0),\label{eq:step_phi-}\\
  &\Phi^\mathrm{step}_+(z)=\frac{1}{32}\left[2(1+z)^2\ln(1+z)-3z^2-2z\right], \quad(z>0), \label{eq:step_phi+}
\end{align}
where the superscripts indicate that these are for the step initial condition.
By the Legendre transform, \eqref{eq:Legendre_phi2mu}, we obtain
\begin{align}
    &\mu^\mathrm{step}_-(k)=-k-\frac{e^{-2k}-1}{e^{-2k}+1}, \quad (k<0),  \label{eq:step_mu-} \\
    &\mu^\mathrm{step}_+(k)=-k+\frac{1}{32}\left[e^{2\left(1+W\left(\frac{8k-1}{e}\right)\right)}\right. \notag \\
    &\hspace{40pt}\left. +2(8k-1)e^{1+W\left(\frac{8k-1}{e}\right)}+1\right], \quad (k>0), \label{eq:step_mu+}
\end{align}
with the Lambert W function $W(\cdot)$ \cite{corless1996}.
Substituting it to \eqref{eq:lambda_2regimes}, we obtain the following predictions on the behavior of $\lambda(k,t)$ for large $t$:
\begin{equation}
    \lambda(k,t) \simeq \begin{cases} \mu^\mathrm{step}_-(k)t, & (k<0), \\ \frac{8}{3}k^{3/2}t^{1/2}, & (k>0). \end{cases}  \label{eq:step_lambda_ld}
\end{equation}
Here, the expression for $k>0$ is obtained by taking the small $k$ limit of \eqref{eq:step_mu+}, $\mu^\mathrm{step}_+(k) \simeq \frac{8}{3}k^{3/2}$.

If $t$ is not large enough, the averaging of \eqref{eq:lambdaDef} may still be dominated by typical fluctuations.
For the step initial condition, typical flucutations are given by \cite{Johansson-CMP2000}
\begin{multline}
    P(\delta h, t) \simeq \frac{\rho^{1/3}(1-\rho)^{1/3}}{t^{1/3}}{} f_\mathrm{GUE}\left(-\rho^{1/3}(1-\rho)^{1/3}\frac{\delta h}{t^{1/3}}\right), \\ (\delta h \sim \mathcal{O}(t^{1/3})), \label{eq:step_GUE}
\end{multline}
with $f_\mathrm{GUE}(\cdot)$ being the probability density function of the GUE Tracy-Widom distribution \cite{Tracy.Widom-CMP1994}.
Substituting it to \eqref{eq:lambdaDef}, we obtain the following expression of the CGF in this regime:
\begin{equation}
  \lambda(k,t) \simeq -\frac{\expct{\chi_\mathrm{GUE}}}{\rho^{1/3}(1-\rho)^{1/3}}kt^{1/3}, \quad(\text{small $|k|$ and $t$}), \label{eq:step_lambda_typical}
\end{equation}
with $\expct{\chi_\mathrm{GUE}} \approx -1.77$ being the mean value of the GUE Tracy-Widom distribution.

The above theoretical predictions are entirely confirmed by our population dynamics simulations (\figref{fig:step}).
The CGF $\lambda(k,t)$ indeed shows crossover from the typical fluctuation regime [\eqref{eq:step_lambda_typical}] to the large deviation regimes [\eqref{eq:step_lambda_ld}], as demonstrated by the precise agreement between the numerical data (symbols) and the theoretical predictions (lines) in \figref{fig:step}(a)(b) for $k<0$ and (d)(e) for $k>0$.
For $k<0$, the absolute value $|\lambda(k,t)|$ is shown, because $\lambda(k,t)$ is negative for small $|k|$ and $t$ as predicted by \eqref{eq:step_lambda_typical}.
As $\lambda(k,t)$ crossovers from the typical fluctuation regime to the large deviation regime, its sign changes from negative to positive, appearing as a cusp in \figref{fig:step}(a) for $k=-0.774$.

Using the data for large $|k|$ and $t$, we can evaluate the scaled CGF $\mu_\pm(\cdot)$.
For $k<0$, $\mu_-(k)$ is evaluated by $\lambda(k,t)/t$ and found to agree precisely with the exact solution [\eqref{eq:step_mu-}] [\figref{fig:step}(c)], except for the leftmost data point for which the convergence was slow and not reached by the simulation time we adopted ($t=10^4$).
For $k>0$, $\mu_+(k/t)$ is evaluated by $\lambda(k,t)/t^2$ [\figref{fig:step}(f)].
One can in principle access the entire functional form of the scaled CGF by $\mu_+(k/t) \simeq \lambda(k,t)/t^2$ [\eqref{eq:lambda_2regimes}], but in practice, a large clone number $N_\mathrm{cl}$ is required to appropriately evaluate $\lambda(k,t)$ for large $k$ and $t$.
For a given finite $N_\mathrm{cl}$, $\lambda(k,t)$ follows \eqref{eq:lambda_2regimes} only for finite time, then it suddenly drops as we can see in the data for $k=1$ in \figref{fig:step}(d).
This limited our access to $\mu_+(k/t)$ to the small-$k/t$ region only, $\mu_+(k/t) \simeq \frac{8}{3}(k/t)^{3/2}$ [\figref{fig:step}(f)].

Before ending this section, it is worthwhile to remark that the large deviation regimes are connected to the typical fluctuation regime through the tails of the GUE Tracy-Widom distribution.
For the posivite tail ($\delta h > 0$), since $f_\mathrm{GUE}(s) \sim e^{-|s|^3/12}$ for $s \to -\infty$, we have $P(\delta h, t) \sim e^{-\frac{1}{48}t^2(\delta h/t)^3}$, hence $\Phi^\mathrm{step}_+(z) \simeq \frac{1}{48}z^3$ for $z \to +0$ and $\mu^\mathrm{step}_+(k) \simeq \frac{8}{3}k^{3/2}$ for $k \to +0$.
These are consistent with the corresponding asymptotic behavior of \eqsref{eq:step_phi+} and \pref{eq:step_mu+}, respectively, and the latter was demonstrated in \figref{fig:step}(f).
Similarly, for the negative tail ($\delta h < 0$), since $f_\mathrm{GUE}(s) \sim e^{-\frac{4}{3}s^{3/2}}$ for $s \to +\infty$, we have $P(\delta h, t) \sim e^{-\frac{2}{3}t|\delta h/t|^{3/2}}$, hence $\Phi^\mathrm{step}_-(z) \simeq \frac{2}{3}|z|^{3/2}$ for $z \to -0$ and $\mu^\mathrm{step}_-(k) \simeq \frac{1}{3}|k|^3$ for $k \to -0$.
These correspond to the asymptotics of \eqsref{eq:step_phi-} and \pref{eq:step_mu-}.
In contrast to the positive tail, our simulations for the negative tail were able to access both the small-$|k|$ and large-$|k|$ regions of $\mu^\mathrm{step}_-(k)$, $\mu^\mathrm{step}_-(k) \simeq \frac{1}{3}|k|^3$ and $\mu^\mathrm{step}_-(k) \simeq |k|-1$, respectively, as demonstrated in \figref{fig:step}(c).

\section{Flat initial condition} \label{sec:flat}

\begin{figure*}
  \centering
  \includegraphics[keepaspectratio,width=\hsize]{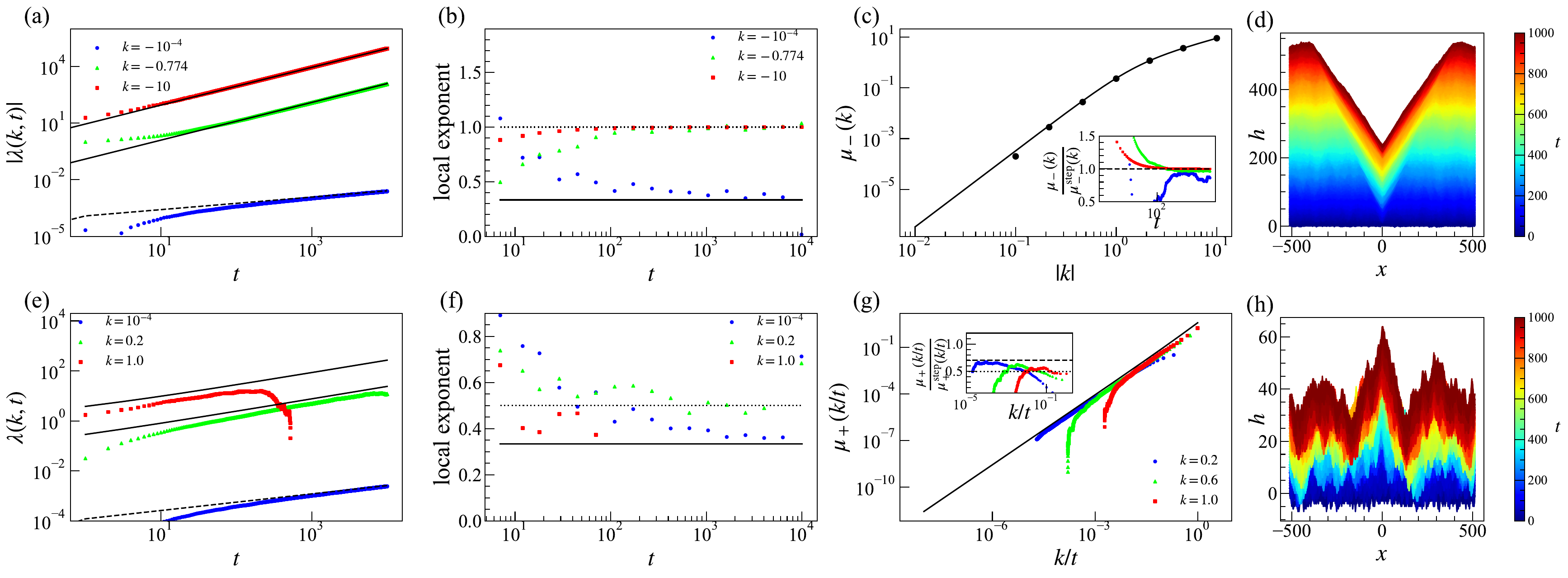}
  \caption{
    Simulation results for the flat initial condition with $N=10000$ (unless otherwise stipulated) and $N_\mathrm{cl} = 10000$.
    (a)(b) The absolute value of CGF $|\lambda(k,t)|$ (a) and its local exponent $\diff{\ln |\lambda(k,t)|}{\ln t}$ (b) for negative bias $k<0$.
    In (a), simulation data for the flat initial condition (symbols) are compared with the exact solution for the step initial condition (solid lines), \eqref{eq:step_mu-}, or the prediction for typical fluctuations for the flat initial condition (dashed line), \eqref{eq:flat_lambda_typical}.
    (c) The scaled CGF $\mu_-(k)$ for the negative tail.
    Simulation data of $\lambda(k,t)/t$ with $t=10000$ for the flat initial condition (symbols) are compared with the exact solution for the step initial condition (line), \eqref{eq:step_mu-}.
    Inset: the ratio of $\lambda(k,t)/t$ for the flat initial condition to the exact solution of $\mu^\mathrm{step}_-(k)$ for the step initial condition, with $k=-0.215$ (blue), $-1$ (green), and $-10$ (red).
    (d) An example of time evolution of interface profiles for $k=-1$, here with $N=1024$. 
    (e)(f) The CGF $\lambda(k,t)$ (e) and its local exponent $\diff{\ln \lambda(k,t)}{\ln t}$ (f) for positive bias $k>0$.
    In (e), simulation data for the flat initial condition (symbols) are compared with the exact solution for the step initial condition (solid lines), \eqref{eq:step_mu+}, or the prediction for typical fluctuations for the flat initial condition (dashed line), \eqref{eq:flat_lambda_typical}.    
    (g) The scaled CGF $\mu_+(k/t)$ for the positive tail.
    Simulation data of $\lambda(k,t)/t^2$ for the flat initial condition (symbols) are plotted against $k/t$ and compared with the exact solution for the step initial condition (line), \eqref{eq:step_mu+}.
    Inset: the ratio of $\lambda(k,t)/t^2$ for the flat initial condition to the exact solution of $\mu^\mathrm{step}_+(k/t)$ for the step initial condition. The dashed and dotted lines are guides for the eyes indicating $1/\sqrt{2}$ and $1/2$, respectively.
    (h) An example of time evolution of interface profiles for $k=1$, here with $N=1024$. 
  }
  \label{fig:flat}
\end{figure*}

For the flat initial condition, to our knowledge, exact solutions for the large deviations are not available in an explicit form in the literature.
Our population dynamics simulations may therefore provide the first report of its large deviation properties.
For typical fluctuations, it is known from exact studies \cite{Borodin.etal-JSP2007} that
\begin{multline}
    \hspace{-15pt} P(\delta h, t) \simeq \frac{\rho^{1/3}(1-\rho)^{1/3}}{t^{1/3}}{} f_\mathrm{GOE}\left(-\rho^{1/3}(1-\rho)^{1/3}\frac{2^{2/3}\delta h}{t^{1/3}}\right), \\ (\delta h \sim \mathcal{O}(t^{1/3})), \label{eq:flat_GOE}
\end{multline}
with $f_\mathrm{GOE}(\cdot)$ being the probability density function of the GOE Tracy-Widom distribution \cite{Tracy.Widom-CMP1996}.
Substituting it to \eqref{eq:lambdaDef}, we obtain the following expression of the CGF in this regime:
\begin{equation}
  \lambda(k,t) \simeq -\frac{\expct{\chi_\mathrm{GOE}}}{2^{2/3}\rho^{1/3}(1-\rho)^{1/3}}kt^{1/3}, \quad(\text{small $|k|$ and $t$}), \label{eq:flat_lambda_typical}
\end{equation}
with $\expct{\chi_\mathrm{GOE}} \approx -1.21$ being the mean value of the GOE Tracy-Widom distribution.
Since $f_\mathrm{GOE}(s) \sim e^{-\frac{2}{3}s^{3/2}}$ for $s \to +\infty$, we have $P(\delta h, t) \sim e^{-\frac{2}{3}t|\delta h/t|^{3/2}}$ in the crossover region to the negative tail ($\delta h < 0$), hence $\Phi^\mathrm{flat}_-(z) \simeq \frac{2}{3}|z|^{3/2}$ for $z \to -0$ and $\mu^\mathrm{flat}_-(k) \simeq \frac{1}{3}|k|^3$ for $k \to -0$, which are identical to those for the step initial condition.
For the potisive tail ($\delta h > 0$), since $f_\mathrm{GOE}(s) \sim e^{-|s|^3/24}$ for $s \to -\infty$, we have $P(\delta h, t) \sim e^{-\frac{1}{24}t^2(\delta h/t)^3}$, hence $\Phi^\mathrm{flat}_+(z) \simeq \frac{1}{24}z^3$ for $z \to +0$ and $\mu^\mathrm{flat}_+(k) \simeq \frac{4\sqrt{2}}{3}k^{3/2}$ for $k \to +0$.
In particular, we have
%This indicates that the scaled CGF of the positive tail differs from that for the step initial condition by factor $2^{-1/2}$ for $k \to +0$, i.e., $\lim_{k \to +0}\frac{\mu_+^\mathrm{flat}(k)}{\mu_+^\mathrm{step}(k)}=2^{-1/2}$.
\begin{equation}
    \lim_{k \to +0}\frac{\mu_+^\mathrm{flat}(k)}{\mu_+^\mathrm{step}(k)}=\frac{1}{\sqrt{2}}.  \label{eq:flat_muratio+}
\end{equation}
Therefore, for the flat case, we expect
\begin{equation}
    \lambda(k,t) \simeq \begin{cases} \mu^\mathrm{flat}_-(k)t, & (k<0), \\ \frac{4\sqrt{2}}{3}k^{3/2}t^{1/2}, & (k>0). \end{cases}  \label{eq:flat_lambda_ld}
\end{equation}

Figure~\ref{fig:flat} shows the numerical results of our population dynamics simulations for the flat case.
Similarly to the step initial condition, the CGF $\lambda(k,t)$ shows crossover from the typical fluctuation regime [\eqref{eq:flat_lambda_typical}] to the large deviation regimes [\eqref{eq:flat_lambda_ld}] [\figref{fig:flat}(a)(b)(e)(f)].
Strikingly, the data indicate that the scaled CGF for the negative tail is identical to that for the step initial condition [\figref{fig:flat}(c)], not only for the $k \to -0$ limit as suggested by the Tracy-Widom tail but for all $k$, i.e., 
\begin{equation}
    \mu^\mathrm{flat}_-(k) = \mu^\mathrm{step}_-(k)  \label{eq:flat_muratio-}
\end{equation}
with $\mu^\mathrm{step}_-(k)$ given by \eqref{eq:step_mu-}.
This may be surprising at first glance, but some hint is given by the interface profile for $k<0$ [\figref{fig:flat}(d)], which develops a wedge very similar to that for the step initial condition [\figref{fig:profile}(a)].
This is indeed reasonable because, for $k<0$, particles are prevented from hopping from $x<0$ to $x>0$, ending up jamming leftward from $x=0$, just like in the case of the step initial condition.
Regarding the positive tail, we confirm $\mu_+(k/t) \sim (k/t)^{3/2}$ with the same exponent value as for the step case [\figref{fig:flat}(g)], but the proportionality constant seems to differ by a factor of $1/\sqrt{2}$ for small $k/t$ (see inset), as predicted in \eqref{eq:flat_muratio+}. 
Our data for different $k$ indicate that the ratio $\mu_+(k/t)/\mu^\mathrm{step}_+(k/t)$ is not a constant [\figref{fig:flat}(g) inset].
Therefore, unlike the negative tail, the positive tail is different between the flat and step cases.
This is reasonable from the viewpoint of interface profiles too, which maintain a globally flat shape for $k>0$ [\figref{fig:flat}(h)] while it is curved in the case of the step initial condition [\figref{fig:profile}(c)].

\section{Stationary initial condition}  \label{sec:stationary}

\begin{figure*}
  \centering
  \includegraphics[keepaspectratio,width=\hsize]{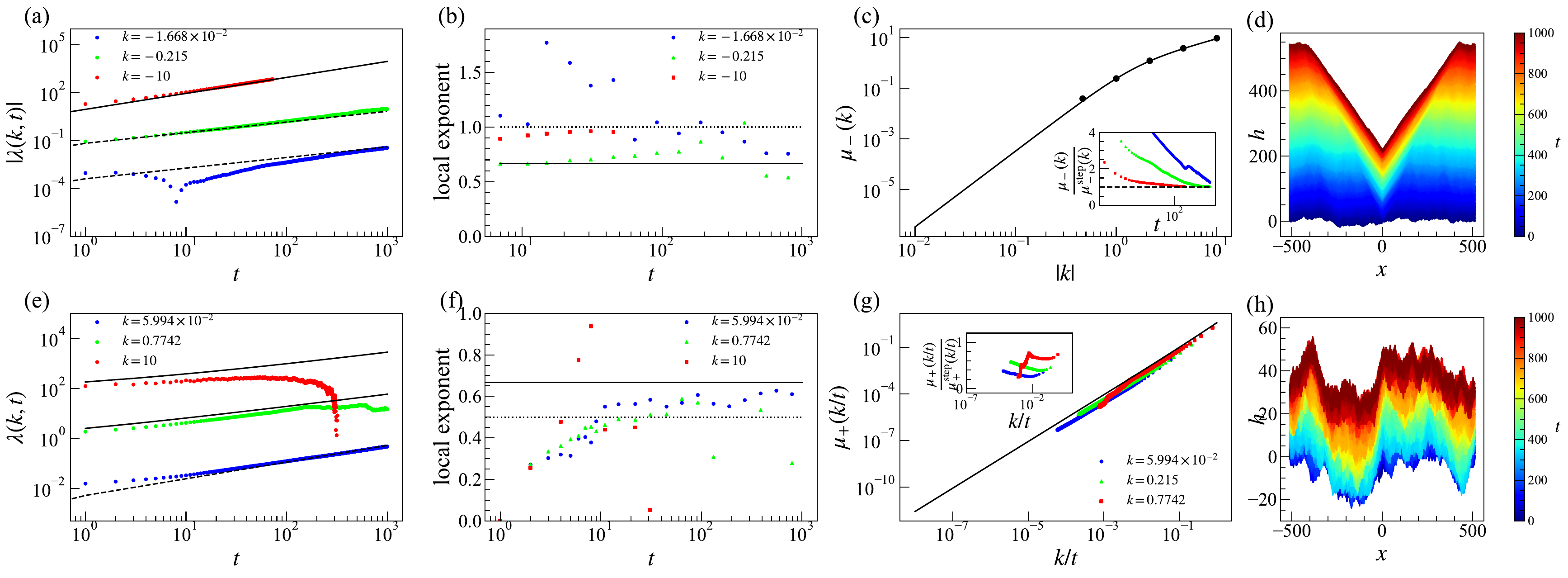}
  \caption{
    Simulation results for the stationary initial condition with $N=10000$ and $N_\mathrm{cl} = 1000$ from $1000$ realizations (unless otherwise stipulated). 
    (a)(b) The absolute value of CGF $|\lambda(k,t)|$ (a) and its local exponent $\diff{\ln |\lambda(k,t)|}{\ln t}$ (b) for negative bias $k<0$.
    In (a), simulation data for the stationary initial condition (symbols) are compared with the exact solution for the step initial condition (solid lines), \eqref{eq:step_mu-}, or the prediction for typical fluctuations for the stationary initial condition (dashed line), \eqref{eq:stat_lambda_typical}.
    (c) The scaled CGF $\mu_-(k)$ for the negative tail.
    Simulation data of $\lambda(k,t)/t$ for the stationary initial condition (symbols) are compared with the exact solution for the step initial condition (line), \eqref{eq:step_mu-}.
    The values of $t$ used are $1000, 1000, 575, 184, 75$ from the leftmost to the rightmost datapoints. 
    Inset: the ratio of $\lambda(k,t)/t$ for the stationary initial condition to the exact solution of $\mu^\mathrm{step}_-(k)$ for the step initial condition, with $k=-0.464$ (blue), $-1$ (green), and $-4.64$ (red).
    (d) An example of time evolution of interface profiles for $k=-1$, here with $N=1024$ and $N_\mathrm{cl} = 10000$.
    (e)(f) The CGF $\lambda(k,t)$ (e) and its local exponent $\diff{\ln \lambda(k,t)}{\ln t}$ (f) for positive bias $k>0$.
    In (e), simulation data for the stationary initial condition (symbols) are compared with the exact solution for the step initial condition (solid lines), \eqref{eq:step_mu+}, or the prediction for typical fluctuations for the stationary initial condition (dashed line), \eqref{eq:flat_lambda_typical}.    
    Inset: the scaled CGF $\mu_+(k/t)$ for the positive tail.
    Simulation data of $\lambda(k,t)/t^2$ for the stationary initial condition (symbols) are plotted against $k/t$ and compared with the exact solution for the step initial condition (line), \eqref{eq:step_mu+}.
    Inset: the ratio of $\lambda(k,t)/t^2$ for the stationary initial condition to the exact solution of $\mu^\mathrm{step}_+(k/t)$ for the step initial condition.
    %The dashed and dotted lines are guides for the eyes indicating $XXX$ and $XXX$, respectively.
    (h) An example of time evolution of interface profiles for $k=1$, here with $N=1024$ and $N_\mathrm{cl} = 10000$.
  }
  \label{fig:stat}
\end{figure*}

Exact solutions to the large deviations are not available for the stationary initial condition either.
Typical fluctuations were exactly solved \cite{Ferrari.Spohn-CMP2006} and known to be
\begin{multline}
    P(\delta h, t) \simeq \frac{\rho^{1/3}(1-\rho)^{1/3}}{t^{1/3}}{} f_\mathrm{BR}\left(-\rho^{1/3}(1-\rho)^{1/3}\frac{\delta h}{t^{1/3}}\right), \\ (\delta h \sim \mathcal{O}(t^{1/3})), \label{eq:stat_BR}
\end{multline}
with $f_\mathrm{BR}(\cdot)$ being the probability density function of the Baik-Rains distribution \cite{Baik.Rains-JSP2000}.
Unlike the Tracy-Widom distributions, the Baik-Rains distribution has vanishing mean \cite{Baik.Rains-JSP2000}, and this results in a scaling of $\lambda(k,t)$ different from that for the step and flat cases.
Specifically, using $\expct{e^{k\delta h}} \simeq 1+\frac{1}{2}\expct{\delta h^2}k^2$ in \eqref{eq:lambdaDef}, we obtain
\begin{equation}
  \lambda(k,t) \simeq -\frac{\expct{\chi_\mathrm{BR}^2}}{2\rho^{2/3}(1-\rho)^{2/3}}k^2t^{2/3}, \quad(\text{small $|k|$ and $t$}), \label{eq:stat_lambda_typical}
\end{equation}
with $\expct{\chi_\mathrm{BR}^2} \approx 1.15$ being the variance of the Baik-Rains distribution.
It is known \cite{Prahofer.Spohn-PRL2000,Baik.Rains-JSP2000,Baik.etal-CMP2008,Ferrari.Veto-ECP2021} that $f_\mathrm{BR}(s) \sim e^{-\frac{2}{3}s^{3/2}}$ for $s \to +\infty$, leading to $P(\delta h, t) \sim e^{-\frac{1}{3}t|\delta h/t|^{3/2}}$ in the crossover region to the negative tail ($\delta h < 0$), hence $\Phi^\mathrm{stat}_-(z) \simeq \frac{1}{3}|z|^{3/2}$ for $z \to -0$ and $\mu^\mathrm{stat}_-(k) \simeq \frac{4}{3}|k|^3$ for $k \to -0$.
Comparing with the value for the step and flat initial conditions, we obtain
\begin{equation}
    \lim_{k \to -0}\frac{\mu_-^\mathrm{stat}(k)}{\mu_-^\mathrm{step}(k)}=4.  \label{eq:stat_muratio-}
\end{equation}
For the potisive tail ($\delta h > 0$), $f_\mathrm{BR}(s) \sim e^{-|s|^3/12}$ for $s \to -\infty$ \cite{Prahofer.Spohn-PRL2000,Baik.Rains-JSP2000,Baik.etal-CMP2008}, leading to $P(\delta h, t) \sim e^{-\frac{1}{48}t^2(\delta h/t)^3}$, hence $\Phi^\mathrm{stat}_+(z) \simeq \frac{1}{48}z^3$ for $z \to +0$ and $\mu^\mathrm{stat}_+(k) \simeq \frac{8}{3}k^{3/2}$ for $k \to +0$.
We thereby obtain
%This indicates that the scaled CGF of the positive tail differs from that for the step initial condition by factor $2^{-1/2}$ for $k \to +0$, i.e., $\lim_{k \to +0}\frac{\mu_+^\mathrm{flat}(k)}{\mu_+^\mathrm{step}(k)}=2^{-1/2}$.
\begin{equation}
    \lim_{k \to +0}\frac{\mu_+^\mathrm{stat}(k)}{\mu_+^\mathrm{step}(k)}=1.  \label{eq:stat_muratio+}
\end{equation}
Therefore, for the stationary case, we expect
\begin{equation}
    \lambda(k,t) \simeq \begin{cases} \mu^\mathrm{stat}_-(k)t, & (k<0), \\ \frac{8}{3}k^{3/2}t^{1/2}, & (k>0). \end{cases}  \label{eq:stat_lambda_ld}
\end{equation}

Figure~\ref{fig:stat} shows the numerical results of our population dynamics simulations for the stationary case.
Since the stationary initial condition is stochastic, here we repeated simulations from different initial realizations $1000$ times and took average of $\lambda(k,t)$ over all realizations.
Then the CGF $\lambda(k,t)$ shows crossover from the typical fluctuation regime [\eqref{eq:stat_lambda_typical}], $\lambda(k,t) \sim t^{2/3}$ which is characteristic of the stationary case, to the large deviation regimes [\eqref{eq:stat_lambda_ld}] [\figref{fig:stat}(a)(b)(e)(f)].
Due to the larger exponent value of the typical fluctuation regime, we were able to reach the large deviation regimes only for relatively large values of $|k|$, compared to the step and flat cases.
This prevented us from testing \eqref{eq:stat_muratio-} for $k \to -0$.
For relatively large $|k|$ for which we could reach the large deviation regime, our data indicate $\mu^\mathrm{stat}_-(k) \simeq \mu^\mathrm{step}_-(k)$ [\figref{fig:stat}(c)], that is, the scaled CGF $\mu_-(k)$ remains identical to that for the step initial condition.
%for the range of $k$ we could investigate, the scaled CGF $\mu_-(k)$ remained identical to that for the step initial condition [\figref{fig:stat}(c)].
%For relatively large $|k|$, our data indicate $\mu^\mathrm{stat}_-(k) \simeq \mu^\mathrm{step}_-(k)$.
Similarly to the flat case, this coincidence can be interpreted as a result of the spontaneous formation of wedge interface profiles for $k<0$ [\figref{fig:flat}(d)].
Regarding the positive tail, we confirm $\mu_+(k/t) \sim (k/t)^{3/2}$ [\figref{fig:stat}(g)], but we were unable to demonstrate the data collapse with different $k$ (inset), most likely due to the limited time range $t \leq 10^3$ we had to work with for repeating many realizations.
It is worthwhile to accomplish this computational task in future, in view of the conjecture for the KPZ equation \cite{Corwin.etal-PRL2018,2note} that $\Phi_\pm^{(\mp)}(z)$, and equivalently $\mu_\pm^{(\mp)}(k)$ (taking the signs in the same order), are identical between the step and stationary initial conditions.

\section{Concluding remarks} \label{sec:conclusion}

In the present work, we carried out population dynamics simulations of TASEP and studied large deviations of the local height (equivalently the integrated current) for the step, flat, and stationary initial conditions, which correspond to the three representative subclasses of the 1D KPZ class.
For the step initial condition, our numerical results agreed with the exact theoretical predictions available in the literature (\figref{fig:step}), for all regimes: positive and negative large deviations [\eqsref{eq:step_mu-}-\pref{eq:step_lambda_ld}] and typical fluctuations [\eqref{eq:step_lambda_typical}]. 
This demonstrates the validity of our numerical scheme.
For the flat initial condition, our numerical results indicate that the scaled CGF $\mu^\mathrm{flat}_-(k)$ for negative large deviations is identical to that for the step initial condition [\eqref{eq:step_mu-}] for all $k<0$, while theory guarantees it only for $k \to -0$.
This may be surprising in view of the initial condition dependence of the KPZ class, but we argue that this may result from the sponetaneously formed wedge geometry of interface profiles in the case of negative bias [\figref{fig:flat}(d)].
By contrast, positive large deviations turned out to be different between the flat and step cases [\figref{fig:flat}(g) inset], which is again understandable from the shape of the interfaces.
Finally, for the stationary initial condition, we first showed theoretically that the CGF $\lambda(k,t)$ in the typical fluctuation regime shows a characterstic power law $\lambda(k,t) \sim t^{2/3}$ [\eqref{eq:stat_lambda_typical}] different from $\lambda(k,t) \sim t^{1/3}$ for the other cases, due to the vanishing mean of the Baik-Rains distribution.
This scaling was indeed confirmed by our numerical data [\figref{fig:stat}(a)(b)(e)(f)].
The data for the large deviation regimes also showed the CGF exponent predicted by theory, though the functional forms of the scaled CGF $\mu^\mathrm{stat}_\pm(k)$ are yet to be determined, due to the much higher computational cost needed for this case.

From broader perspectives, our results demonstrate that the population dymamics method can be reliably used to investigate large deviations of systems in the KPZ class.
Comparing with the other kind of importance sampling method that modifies the time evolution rule, as implemented previously \cite{Hartmann.etal-EL2018,Hartmann.etal-PRE2020}, the population dynamics method may be heavier in computational cost but is more versatile, as it does not require modifying nor even knowing the time evolution rule.
Therefore, it can in principle be implemented even experimentally.
In this context, the laser holographic technique, which was used to generate arbitrary initial conditions in the liquid-crystal KPZ experiments \cite{Fukai.Takeuchi-PRL2017,Fukai.Takeuchi-PRL2020,Iwatsuka.etal-PRL2020}, is particularly interesting as it can also be used to clone an observed interface.

\begin{acknowledgments}
%\paragraph{Acknowledgments.} 

We thank P. Le Doussal for fruitful discussions.
This work is supported in part by JSPS KAKENHI Grant Numbers JP20H01826, JP23K17664, JP24K00593 and the JSPS Core-to-Core Program ``Advanced core-to-core network for the physics of self-organizing active matter'' (JPJSCCA20230002). 

\end{acknowledgments}

%\begin{thebibliography}{99}
  
%\end{thebibliography}
\bibliography{tasep.bib}

\end{document}